\documentclass[aps,prb,superscriptaddress,preprint]{revtex4}  
\usepackage[reqno]{amsmath} 
\usepackage{graphicx} 
\usepackage{makeidx} 

\begin{document} 

\title{Valence holes as Luttinger spinor based qubits in quantum dots} 

\author{Chang-Yu Hsieh}
\affiliation{Quantum Theory Group, 
Institute for Microstructural Sciences, 
National Research Council, Ottawa, ON, Canada K1A 0R6} 
\affiliation{Department of Physics, 
University of Ottawa, Ottawa, ON, Canada, K1N 6N5} 

\author{Ross Cheriton}
\affiliation{Quantum Theory Group, 
Institute for Microstructural Sciences, 
National Research Council, Ottawa, ON, Canada K1A 0R6} 
\affiliation{Department of Physics, 
University of Ottawa, Ottawa, ON, Canada, K1N 6N5} 

\author{Marek Korkusinski}
\affiliation{Quantum Theory Group, 
Institute for Microstructural Sciences, 
National Research Council, Ottawa, ON, Canada K1A 0R6} 

\author{Pawel Hawrylak}
\affiliation{Quantum Theory Group, 
Institute for Microstructural Sciences, 
National Research Council, Ottawa, ON, Canada K1A 0R6} 
\affiliation{Department of Physics, 
University of Ottawa, Ottawa, ON, Canada, K1N 6N5} 

\begin{abstract} 
 We present a theory of valence holes as Luttinger spinor based qubits in p-doped self-assembled quantum dots within the 4-band $k\cdot p$ formalism.   The two qubit levels are identified with the two chiralities of the doubly degenerate ground state.  We show that single qubit operations can be implemented with static magnetic field applied along the $z$ axis (growth direction) for $\hat{\sigma}_z$ operation and with magnetic field in the  quantum dot plane, $x$ direction, for $\hat{\sigma}_x$ operation.  The coupling of two dots and hence the double qubit operations are shown to be sensitive to the orientation of the two quantum dots.  For vertical qubit arrays, there exists an optimal qubit separation suitable for the voltage control of qubit-qubit interactions.
\end{abstract} 

\maketitle 

\section{Introduction}
There is currently interest in using single electron spins in quantum dots (QDs) for quantum information processing (QIP) applications\cite{brum_hawrylak_sm1997,loss_divincenzo_pra1998,burkard_loss_prb1999,hanson_kouwenhoven_rmp2007,   sachrajda_hawrylak_book2003,korkusinski_hawrylak_book2008}.  In $III-V$ semiconductor quantum dots, such as $GaAs$, spins of conduction band electrons couple to nuclear spins of the host material and suffer decoherence due to hyperfine interactions.  However, valence hole states are built from atomic $p$-type orbitals, which are expected to minimize the hyperfine interaction between the hole and surrounding nuclear spins \cite{gerardot_brunner_nat2007}. The spin of a valence hole as a qubit  is expected to have longer coherence time. 
A theory of heavy holes as qubits has been developed in the single band approximation\cite{gerardot_brunner_nat2007, bulaev_loss_prl2007, bulaev_loss_prl2005}. The justification for the use of the 
heavy hole single band approximation is that strain and QD confinement suppress the coupling between heavy hole (HH) and light hole (LH) bands. Under such an assumption, the qubit levels are defined by  the $J_z=+3/2$ and $J_z= -3/2$ HH states.  
 
 In such a simple model, the exchange coupling $J$ needed to generate entanglement between spins of two holes is proportional to $t^2/U$, where $t$ is the tunneling strength of the hole and $U$ is the on-site Coulomb energy.  However, recent  theoretical \cite{climente_korkusinski_prb2008} and experimental \cite{doty_climente_prl2009}  work show that in two vertically coupled disk-like QDs the spin orbit (SO) coupling between HH and LH bands changes the sign of the effective tunneling matrix element $t$  as a function of inter-dot distance.  If the inter-dot distance is smaller than a critical value, the hole state is mainly symmetric.  However, if the inter-dot distance passes the critical value, the hole state is mainly antisymmetric.  These results suggest that a description of an array of hole-based qubits requires taking into account the hole tunneling between dots with SO coupling between HH and LH bands. The theory of valence holes  in quantum dots with strong SO coupling has been
developed already \cite{pedersen_chang_prb1996, darnhofer_rossier_prb1992, brodio_cros_prb1992, sercel_vahala_prb1990,  rego_hawrylak_prb1997, kyrychenko_kossut_prb2004}. In this theory  hole spin is strongly coupled to orbital motion and valence hole states are treated as Luttinger spinors. The two HH ground states are replaced by two Luttinger spinors \cite{rego_hawrylak_prb1997} with different chiralities.
 
\begin{figure}
\centering
\includegraphics[width=0.45\textwidth]{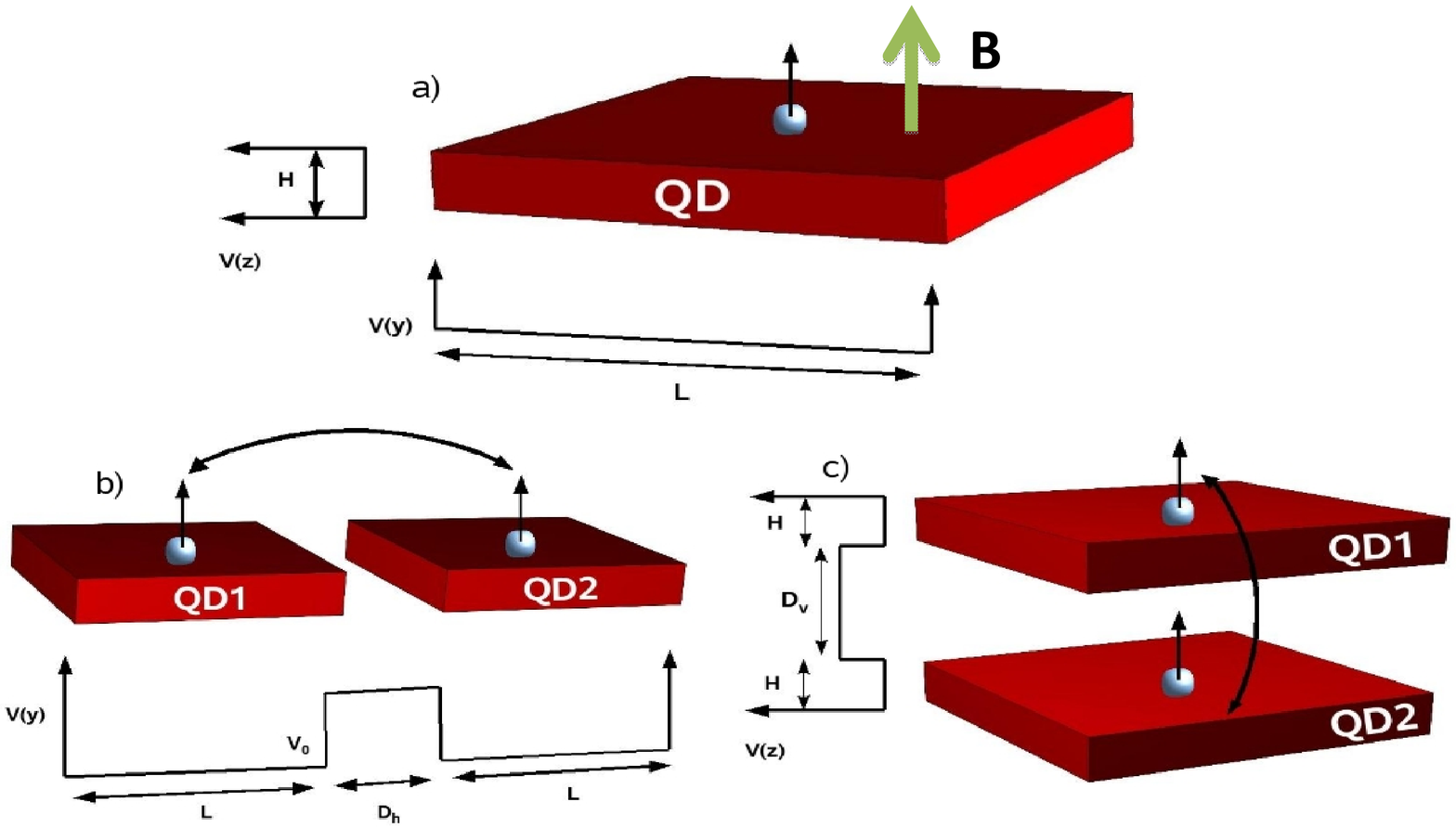}  
\caption{Schematics of (a) square vertically coupled quantum dots (VCQDs) of side lengths L=20nm and height H=2nm separated by a potential barrier $V_0$ of width $D_v$ along the vertical direction.  (b) Square laterally coupled quantum dots (LCQDs) of same side lengths L, height H separated by a potential barrier $V_0$ of width $D_h$ along the horizontal direction.}
\label{fig:schem}
\end{figure}

In this work, we develop a theory of qubits based on the chirality of the valence hole-based Luttinger spinors.
We are particularly interested in square-like self-assembled quantum dots grown on nanotemplates \cite{reimer_mckinnon_phyE2007} , which due to scalable architecture may lead
to  quantum information processing devices. We start with a single hole confined in an isolated QD.  We explicitly define the qubit as two chirality states of Luttinger spinor and show how to perform  single qubit operations. Our theory reproduces results obtained earlier by Kyrychenko and Kossut \cite{kyrychenko_kossut_prb2004}. Next, we investigate the tunneling of a valence hole in two orientations of two coupled square QDs as illustrated in Fig \ref{fig:schem}.  The tunneling barriers for vertically coupled quantum dots (VCQDs) and laterally coupled quantum dots (LCQDs) are modelled as finite potential wells, as shown in Fig \ref{fig:schem}.  The tunnel barrier strength is characterized by the band-offset between the dot and the barrier material. For VCQDs, we verify the results \cite{doty_climente_prl2009} obtained for disk quantum dots showing the reversal of ground state character from symmetric to antisymmetric as a function of inter-dot distance. Close to the reversal, the vanishing of tunneling of a  Luttinger spinor hole  in VCQDs allows the benefit of strong confinement defined by growth with the possibility to control tunneling and hence double qubit operations using additional metallic gates.

\section{Model}
    Following Ref. \onlinecite{rego_hawrylak_prb1997} , we expand the wavefunction of a valence hole confined to a nanostructure defined by confining potential $V(\mathbf{r})$   in terms of  $J_z = 3/2, -1/2, 1/2, -3/2$ Bloch basis functions. The 4-band Luttinger Kohn (LK) Hamiltonian \cite{luttinger_kohn_pr1955}  reads,    
\begin{equation}
\label{eq:LKmatrix}
\hat{H}_{LK}= \left( \begin{array}{cccc}
\hat{P}^+ & \hat{R} & -\hat{S} & 0\\
\hat{R}^* & \hat{P}^- & 0 & \hat{S}\\
-\hat{S}^* & 0 & \hat{P}^- & \hat{R} \\
0 & \hat{S}^* & \hat{R}^* & \hat{P}^+\end{array}\right) 
+V(\mathbf{r}) \mathbf{I}
+\kappa \Omega \mathbf{J} \cdot \mathbf{\hat{B}} 
,\end{equation}
where $\mathbf{I}$ is the identity matrix, $\hbar$ is Planck's constant, $\Omega = \hbar e B / m_0 c$ is the cyclotron energy, and $\kappa$ is a material parameter.  $V(\mathbf{r})$ represents a 3D infinite potential well in case of a single QD in this study.  The operator $\mathbf{J}$ is the angular momentum operator for spin $3/2$ particle defined in Ref.~\onlinecite{luttinger_kohn_pr1955}.  The operators in Eq.(\ref{eq:LKmatrix}) are defined as follows
\begin{eqnarray}
\label{eq:LKoperators}
\hat{P}^+&=&\frac{\hbar^2}{2m_0}[(\gamma_1+\gamma_2)({\Pi_x}^2+{\Pi_y}^2)+(\gamma_1-2\gamma_2){\Pi_z}^2],\nonumber \\
\hat{P}^-&=&\frac{\hbar^2}{2m_0}[(\gamma_1-\gamma_2)({\Pi_x}^2+{\Pi_y}^2)+(\gamma_1+2\gamma_2){\Pi_z}^2],\nonumber \\
\hat{R} &=&\frac{\hbar^2}{2m_0}(-\sqrt{3})\gamma_{23} {\Pi_-}^2, \nonumber \\
\hat{S} &=&\frac{\hbar^2}{2m_0}(2\sqrt{3})\gamma_3 {\Pi_-}{\Pi_z},
\end{eqnarray}
where $\gamma_{1,2,3}$ represent the Luttinger material parameters, $\gamma_{23}=(\gamma_{2}+\gamma_{3})/2$, $\Pi_a = k_a -(e/c\hbar)\left(\mathbf{A}\right)_a$, $k_{a}=-i\frac{\partial}{\partial{a}}$,  $a = x,y,z$,  and $k_{-}=k_x-{ik_y}$.   For this study, we choose InGaAs/GaAs QDs.  The InGaAs Luttinger material parameters are $\gamma_1=11.01$, $\gamma_2=4.18$, and $\gamma_3=4.84$.  The subscript 0 means no external field in this study.

The LK Hamiltonian in Eq.(\ref{eq:LKmatrix}) exhibits several symmetries that we use to characterize the Luttinger spinors (the eigenstates of the LK Hamiltonian). The confining potential along the $z$ direction is symmetric with respect to the origin defined in the middle of the structure; yet, due to spin-orbit coupling between different bands, our system does not have definite parity symmetry along the $z$ direction but the time reversal symmetry of the system demands the two HH bands have opposite parity.  The same also holds for LH bands.  This allows us to define the chirality operator \cite{rego_hawrylak_prb1997,climente_korkusinski_prb2008} which reads, 
\begin{equation}
\label{eq:chirality}
\hat{\chi}_{z}= \left( \begin{array}{cccc}
\hat{i}_{z}  & 0 & 0 & 0\\
0  & \hat{i}_{z} & 0 & 0\\
0  & 0 & -\hat{i}_{z}& 0 \\
0  & 0 & 0 & -\hat{i}_{z} \end{array}\right),
\end{equation}
where $\hat{i}_{z}$ is the inversion operator with respect to the $z$ variable.  The $\hat{\chi}_z$ operator has 2 eigenvalues, which we denote by $\uparrow$, and $\downarrow$.  In the case of $\chi_z=\uparrow$, the Luttinger spinors have even parity with respect to $z$ for the first two components of the spinor and odd parity with respect to $z$ for the last two components of the spinor.  In the case of $\chi_z=\downarrow$, the parity pattern is reversed between the first two and last two components of the spinor.  Furthermore, square QDs also have parity symmetry in the $x-y$ plane, and we may define another parity operator $\hat{\chi}_{xy}$ analogous to the $\hat{\chi}_z$ by replacing $\hat{i}_{z}$ with $\hat{i}_{x}\hat{i}_{y}$ on the diagonal terms in Eq.(\ref{eq:chirality}) where $\hat{i}_{x}$ and $\hat{i}_{y}$ are inversion operators along the $x$ and the $y$ direction respectively. Arguments for the $\hat{\chi}_z$ operator also apply to the $\hat{\chi}_{xy}$ operator.  The only difference is that the $\hat{R}$ operator in Eq.(\ref{eq:LKoperators}) involves $(k_{-})^2$ term which prevents further separation of $x$ and $y$ variables.   Again, $\hat{\chi}_{xy}$ has 2 eigenvalues denoted by $\chi_{xy} = +1 $ and $\chi_{xy}=-1$.  For $\chi_{xy}=+1$, the first two components of the spinor must have the same parity with respect to $x$ and $y$, and the last two components of the spinor must have opposite parity with respect to $x$ and $y$.  For $\chi_{xy}=-1$, the Luttinger spinors again have the parity pattern switched between the first two and last two components.  

The 4-band valence hole spinors in a square QD have well defined structures.  The Bloch function part of the Luttinger spinors is determined by the lattice symmetries; while the envelope function part of Luttinger spinors is obtained from Eq.(\ref{eq:LKmatrix}).  In our analysis, we expand the envelope function of the Luttinger spinors in eigenstates of the 3D infinite potential well of the size of our computational box.  As mentioned, the chirality and the $x-y$ parity symmetries allow us to label Luttinger spinors by $\chi_{z}$, $\chi_{xy}$, and $N$, which index the $N^{th}$ eigenstate of subspace $\chi_z$ and $\chi_{xy}$.  For instance, the $N^{th}$ state of the subspace $\chi_{z}=\uparrow$ and $\chi_{xy}=+1$ reads
\begin{eqnarray}
\label{eq:spinor}
\vert \uparrow, +1, N \rangle  = 
\sum_{\substack{n+m=2p-1\\n'+m'=2p}}
 \left( \begin{array}{c}
  C_{nm1}^{\uparrow,+1,N, 3/2}\xi_{n}(x)\xi_{m}(y)\cos\left(\frac{\pi z}{W_z}\right) \vert \frac{3}{2}\rangle \\
  C_{nm1}^{\uparrow,+1,N, -1/2}\xi_{n}(x)\xi_{m}(y)\cos\left(\frac{\pi z}{W_z}\right) \vert -\frac{1}{2}\rangle \\
  C_{n'm'2}^{\uparrow,+1,N,1/2}\xi_{n'}(x)\xi_{m'}(y)\sin\left(\frac{2 \pi z}{W_z}\right) \vert \frac{1}{2}\rangle \\
  C_{n'm'2}^{\uparrow,+1,N,-3/2}\xi_{n'}(x)\xi_{m'}(y)\sin\left(\frac{2 \pi z}{W_z}\right) \vert -\frac{3}{2}\rangle \\
\end{array}\right) ,
\end{eqnarray}
%
where $p=1,2,3,...$ are positive integers. $C_{nml}^{\chi_{z},\chi_{xy},N,J_z}$ represents the coefficient for the envelope function of a specific component $\vert J_z \rangle$ of Luttinger spinors.  In our study, we assume that the confinement along the $z$ direction is strong, so we only need to consider the first 2 eigenfunctions of the 1D infinite potential well in this direction.  $\xi_{s_r}(r)=\sqrt{\frac{2}{W_r}}\cos\left(\frac{s_r\pi r}{W_r}\right)$ if $s_r$ is odd and $\sqrt{\frac{2}{W_r}}\sin\left(\frac{s_r\pi r}{W_r}\right)$ if $s_r$ is even. $W_r$ is the computation box length along the $r$ direction.  $ \vert J_z = \pm \frac{3}{2}, \pm\frac{1}{2}\rangle$  represents the Bloch functions \cite{luttinger_kohn_pr1955} of zincblende structure with the total angular momentum projection $J_z$.   The Luttinger spinors in other subspaces have distinguishable parity patterns derived from Eq.(\ref{eq:spinor}) by switching either the  parity pattern with respect to the $x-y$ variable or the $z$ variable between first two and last two components of the spinor.  

From here onwards,  the Luttinger spinor is written with the components corresponding to the Bloch functions  in the order $J_z = 3/2, -1/2, +1/2, -3/2$ as done in Eq.(\ref{eq:spinor}), and we will no longer explicitly write out the Bloch functions $\vert J_z \rangle$ as we have done above. 

\section{Hole states in single quantum dot}
In this analysis, we take the entire Luttinger spinor as a representation of valence hole state confined in the quantum dots as opposed to adopting the HH single band approximation \cite{gerardot_brunner_nat2007,bulaev_loss_prl2007}.  Due to the time reversal symmetry, we have a doubly degenerate ground state with distinct chirality at zero field.  We encode the qubit with the chirality, $\chi_{z}$, of the ground state Luttinger spinors.  We show that static magnetic fields applied along the $z$ and the $x$ direction indeed act as $\hat{\sigma}_x$ and $\hat{\sigma}_z$ operators respectively in the qubit subspace.  This result assures all the basic ingredients needed to perform arbitrary single qubit rotations. 

\subsection{Hole under $B_z$ field}
By diagonalizing the LK Hamiltonian for a square QD, we find that the 2 ground states are characterized by quantum numbers $(\chi_z = \uparrow, \chi_{xy}=-1)$ and $(\chi_z = \downarrow, \chi_{xy}=+1)$ respectively.  Let us  denote the state with $\vert\chi_z =\uparrow \rangle$ by  $\vert 1 \rangle$, and  the other state with $\vert \chi_z = \downarrow \rangle $ by $\vert 0 \rangle$.  We give an approximate form of the envelope wave functions 
\begin{equation}
	\label{eq:state1inSQD}
	\vert 1 \rangle= 
	\left( \begin{array}{c}
	a\xi_1(x)\xi_1(y)\xi_1(z) \\
	b\xi_2(x)\xi_2(y)\xi_1(z) \\
	\left[c_1\xi_1(x)\xi_2(y)+c_2\xi_2(x)\xi_1(y)\right]\xi_2(z) \\
	\left[d_1\xi_1(x)\xi_2(y)+d_2\xi_2(x)\xi_1(y)\right]\xi_2(z) \\
	\end{array}\right) ,
\end{equation}
where $\xi_1(r) = \sqrt{\frac{2}{W_r}}\cos\left(\frac{\pi r}{W_r}\right)$ and $\xi_2(r) = \sqrt{\frac{2}{W_r}}\sin\left(\frac{2 \pi r}{W_r}\right)$.
$\vert 0 \rangle$ is obtained by applying the time reversal operator to $\vert 1\rangle$.  According to Kramer's degeneracy theorem, $\vert 0\rangle$ will have the same set of envelope functions in Eq.(\ref{eq:state1inSQD}) but complex conjugated and in a reversed order in the spinor.   Due to the strong confinement along the $z$ direction, the ground states of the system will contain a dominant contribution from the HH states.  Therefore, the states $\vert 0 \rangle$ and $\vert 1 \rangle$ correspond to either a $+3/2$ HH or $-3/2$ HH occupying the $(1,1,1)$ orbital of a 3D infinite potential box. 

To analyze the system in the presence of external fields $\mathbf{B}$, we use the gauge $\mathbf{A} = (-B_z y / 2,B_z x / 2, B_x y - B_y x)$.  First, we consider the system under an external field $B_z$.  Fig \ref{fig:bfields}a  shows the energy spectrum of a single square dot charged with one valence hole as a function of $B_z$.  We see that the qubit subspace, the 2 lowest energy levels, are well isolated from the rest of the energy spectrum across a wide range of applied field.  Due to the weak coupling between the qubit subspace and other excited states, we may treat the effects of $B_z$ fields by the first order L\"{o}wdin perturbation theory\cite{lowdin_chemphys1951}, which is simply a projection of the full LK Hamiltonian into the qubit subspace.  The entire LK Hamiltonian is decomposed into 2 parts: unperturbed Hamiltonian at the zero field and additional perturbing Hamiltonian due to the field.  The additional Hamiltonian with external field $B_z$ leads to $\hat{P}^{+(-)}_1$ operators  of the form
\begin{eqnarray}
\label{eq:LKoperatorsSZ}
\hat{P}^+_1  & = & \frac{\hbar^2 }{2m_0}(\gamma_1 + \gamma_2) \left[  (\omega_z)^2 (x^2 + y^2) + \omega_z (xk_y - yk_x) \right],\nonumber \\
\hat{P}^- _1  & = & \frac{\hbar^2}{2m_0} (\gamma_1 - \gamma_2)\left[  (\omega_z)^2 (x^2 + y^2) +\omega_z (xk_y - yk_x) \right], \nonumber \\
\end{eqnarray}
where $\omega_z = \frac{eB_z}{c\hbar}$.
Whether the applied field can break the chirality symmetry $\hat{\chi}_{z}$ and the x-y parity symmetry $\hat{\chi}_{xy}$ depends on the commutation relations between $\hat{P}^{+(-)}_1$ operators and inversion operators $\hat{i}_{z}$ and $\hat{i}_{x}\hat{i}_{y}$ . Since  $\hat{P}^{+(-)}_1$ commute with $\hat{i}_{z}$, the Hamiltonian can not mix the 2 spinors of different chiralities.
Indeed, when we project the Hamiltonian we find the off-diagonal matrix elements $\langle 1 \vert \hat{H}_{LK} \vert 0 \rangle = 0$, and the diagonal matrix elements $\langle 1 \vert \hat{H}_{LK} \vert 1 \rangle $ and $\langle 0 \vert \hat{H}_{LK} \vert 0 \rangle$ differ only by the Zeeman energy.  The diagonal matrix elements read,
\begin{eqnarray}
\label{eq:Hbz}
\langle 1 \vert \hat{H}_{LK} \vert 1 \rangle  & = & \kappa\Omega(|a|^2-|d_1|^2-|d_2|^2)(3/2)+ \nonumber \\
                                                                            &    &\kappa\Omega(-|b|^2+|c_1|^2+|c_2|^2)(1/2), \nonumber \\
\langle 0 \vert \hat{H}_{LK} \vert 0 \rangle  & = & -\langle 1 \vert \hat{H}_{LK} \vert 1 \rangle,
\end{eqnarray}
where the coefficients $a, b, c, d$ are defined in Eq.(\ref{eq:state1inSQD}).  As the ground state of flat QDs has dominant contributions from HH components, typically $|a|^2 \approx 0.9$ is much greater than the magnitude of other coefficients.  In HH single band approximation,  one simply sets $|a| = 1$ and other coefficients to be 0.

\subsection{Hole under $B_x$ field}

\begin{figure}
\centering
\includegraphics[width=0.9\textwidth]{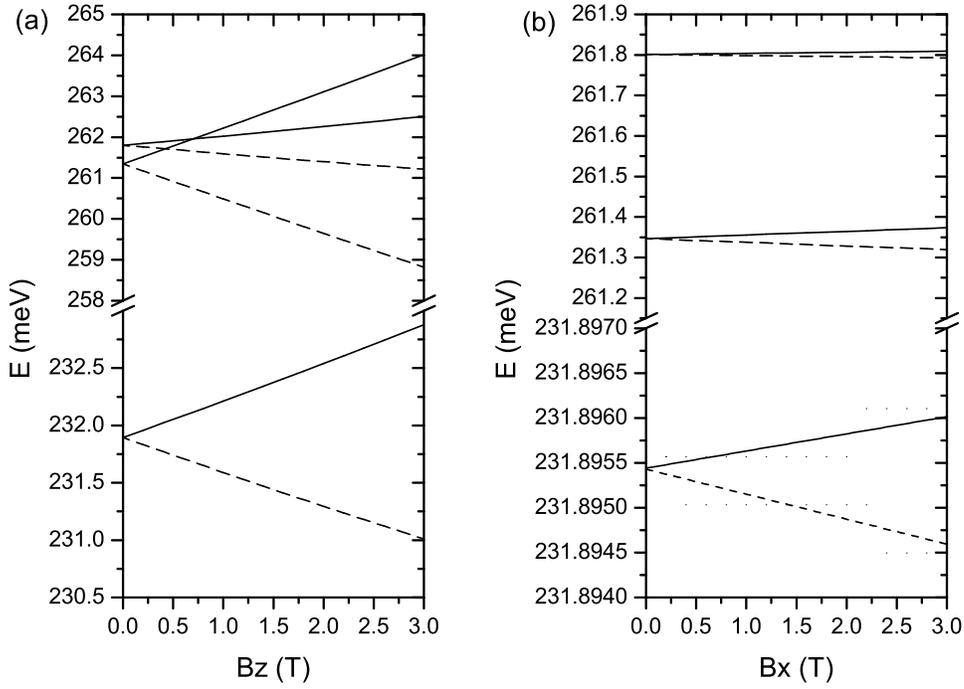}  
\caption{Energy levels of the 6 lowest lying hole states as a function of (a) magnetic field $B_z$ (applied in the z direction) and (b) magnetic field $B_x$ (applied in the x direction).}
\label{fig:bfields}
\end{figure}

Next, we consider the system subject to a constant $B_x$ field.  Fig \ref{fig:bfields}b shows the energy spectrum of a single square QD charged with one valence hole as a function of $B_x$.  The two qubit states are again well isolated from the rest of the energy spectrum across a wide range of $B_x$.   Hence, we repeat the same procedure to derive an effective Hamiltonian for the qubit.  
The vector potential is $\mathbf{A} = (0,0,B_x y)$ and the  $\hat{P}^{+(-)}_1$ operators for the additional LK Hamiltonian with $B_x$ field read
\begin{eqnarray}
\label{eq:LKoperatorsSX}
\hat{P}^+_1  & = & \frac{\hbar^2}{2m_0}(\gamma_1 - 2 \gamma_2) \left[ \left(\omega_x \right)^2 y^2 - 2\omega_x yk_z  \right] ,\nonumber \\
\hat{P}^- _1  & = &  \frac{\hbar^2}{2m_0}(\gamma_1 +2  \gamma_2) \left[ \left(\omega_x \right)^2y^2 - 2\omega_x yk_z \right], \nonumber \\
\end{eqnarray}
where $\omega_x = \frac{eB_x}{c\hbar}$.
We find that the diagonal matrix elements $\langle 1 \vert \hat{H}_{LK} \vert 1 \rangle = \langle 0 \vert \hat{H}_{LK} \vert 0 \rangle$, and the off-diagonal matrix elements are given by
\begin{eqnarray}
\label{eq:Hbx}
\langle 1 \vert \hat{H}_{LK} \vert 0 \rangle  & = & \langle g^{-3/2} \vert  \hat{P}^+_1 \vert g^{3/2} \rangle + \langle g^{3/2} \vert \hat{P}^+_1 \vert g^{-3/2} \rangle  + \nonumber \\
                                                                           &   & \langle g^{-1/2} \vert \hat{P}^-_1 \vert g^{1/2} \rangle + \langle g^{1/2} \vert \hat{P}^-_1 \vert g^{-1/2}  \rangle - \nonumber \\
                                                                           &    & \langle g^{-1/2} \vert \hat{S}^* \vert g^{3/2} \rangle +  \langle g^{3/2} \vert \hat{S}^* \vert g^{-1/2} \rangle - \nonumber \\
                                                                           &   & \langle g^{-3/2} \vert \hat{S} \vert g^{1/2} \rangle + \langle g^{1/2} \vert  \hat{S} \vert g^{-3/2} \rangle, \nonumber \\
& \approx & \Omega_x (\gamma_1 - 2\gamma_2)\frac{W_y}{W_z}Im(d_1^*a).
\end{eqnarray}
where $\Omega_x = \hbar e B_x/m c$. $\vert g^i \rangle$ stands for the envelope function of the $i$ hole states in Eq.(\ref{eq:state1inSQD}), for instance, $\langle \mathbf{r} \vert g^{3/2} \rangle = a\xi_1(x) \xi_1(y) \xi_1(z)$. The matrix elements $\langle g^{-3/2} \vert \hat{P}^+_1 \vert g^{3/2} \rangle$ do not suggest coupling of +3/2 HH and -3/2 HH.  Rather, it actually represents the coupling of a chirality up ($\chi_z=\uparrow$) +3/2 HH with a chirality down ($\chi_z=\downarrow$) +3/2 HH.  Due to time reversal symmetry, the chirality down +3/2 HH must have the same envelope function as the chirality up -3/2 HH.  Similar argument applies to all other matrix elements in Eq.(\ref{eq:Hbx}). We further remark that the off-diagonal matrix element contains pairs of complex conjugates; thus, the matrix element is real-valued. 

The mixing of the qubit states is due to the simultaneous breaking of the parity symmetry in the $x-y$ plane and the inversion symmetry along the $z$ direction; for instance, we refer to the term $yk_z$ in both $\hat{P}^+_1$ and $\hat{P}^-_1$ operators in Eq.(\ref{eq:LKoperatorsSX}) as responsible for the breaking of symmetries. The fact that we may couple the qubit states by static fields is in contradiction to analysis done in the HH single band approximation \cite{bulaev_loss_prl2007}.  In HH single band approximation, the single qubit operation cannot be done by electron spin resonance (ESR) techniques because the magnetic field cannot couple two HH states (in the leading dipole approximation).  More sophisticated techniques such as electric dipole spin resonance (EDSR) are needed.  However,  in our proposal, we define qubit states with the entire Luttinger spinor, which is an admixture of all HHs and LHs, and the mixing of the two qubit states is due to the coupling between the two Luttinger spinors as manifested in Eq.(\ref{eq:Hbx}).  

The $\hat{\sigma}_z$ operator for our proposed Luttinger spinor based qubit is essentially driven by the Zeeman energy; whereas the $\hat{\sigma}_x$ operator is based on the mixing of valence hole components by vector potential $\mathbf{A}$ and SO coupling.
Due to the different mechanisms of how the 2 qubit states are operated on by $B_z$ and $B_x$ field, the effective g factor (for a simple spin in an external field) seems to have much stronger transverse component\cite{kesteren_cosman_prb1990}, $g_{\perp} >> g_{//}$.  Hence it takes more time to perform a spin flip process with the Luttinger spinor based qubit.  We will characterize the single qubit operating time with the $\hbar/\Delta E_{gap}$, where $\Delta E_{gap}$ is the energy difference between ground and first excited state under $B_x$ field.  From Fig \ref{fig:bfields}b, we see the characteristic time is around 1.5 ns when $B_x = 1T$. However, modulating the strain of the host material can relax the QD confinement along the $z$ direction, induce a stronger planar component of the g factors and improve the single qubit operating time.

Finally, we remark that the Luttinger spinor based qubits are susceptible to the same channels of decoherence that are already discussed for valence hole confined in QDs in Refs.~\onlinecite{bulaev_loss_prl2005,woods_reinecke_prb2004, laurent_eble_prl2005,serebrennikov_physlettA2008}.  However, electric field fluctuation in the background of the host material will not affect the Luttinger spinor based qubits.    As the electric field induced dipole transitions, $\langle \psi_j \vert e\mathbf{E}(t)\cdot\mathbf{r} \vert \psi_i \rangle$, conserve the time reversal symmetry, the two qubit states, which are Kramers doublets, will not be mixed.

\section{ Hole states in coupled quantum dots}
We now turn to the analysis of double qubit operations with the Luttinger spinors based qubits.  Our proposal for double qubit operations with the valence holes-based qubits relies on the following assumptions.  First, we consider 2 valence holes localized in two different QDs in the weak tunneling limit and only the on-site Coulomb interaction is taken into account.  Second, we consider that each QD only contains the two relevant qubit states. Under these assumptions, the hole-hole interacting Hamiltonian in second quantization reads \cite{rego_hawrylak_prb1997}
\begin{eqnarray}
\label{eq:2HoleH}
H_{2h}& =& \sum_{j, p} \epsilon_{jp} c_{jp}^\dag c_{jp} + \sum_{j} t \left(c^\dag_{jp'}c_{jp} +c^\dag_{jp}c_{jp'} \right) \nonumber \\
                & &+ \frac{1}{2} \sum_{\substack{j_{1p},j_{2p} \\ j_{3p},j_{4p}\\p}} U_{j_{1p} j_{2p} j_{3p} j_{4p}}c^{\dag}_{j_{1p}}c^{\dag}_{j_{2p}}c_{j_{3p}}c_{j_{4p}},
\end{eqnarray}
where $p$ are indices for QD number 1 and 2, $p \neq p'$, $\epsilon_{jp}$ is the energy of valence hole state $\vert j \rangle$ on the $p-th$ dot, t is the tunneling parameter between QDs, and $U_{j_1j_2j_3j_4}$ is the on-site Coulomb interaction, with the matrix elements given in the appendix.  This Hamiltonian can be greatly simplified because the Coulomb interaction conserves the chirality of 2 holes in the following manner: (a) if $\chi_{z_1} = \chi_{z_2} $ and $\chi_{z_3} = \chi_{z_4}$, $U_{j_1 j_2 j_3 j_4} = U_c$, (b)  if $\chi_{z_1} \neq \chi_{z_2} $ and $\chi_{z_3} \neq \chi_{z_4}$, $U_{j_1 j_2 j_3 j_4} = U_x$, (c) if just one of the valence holes switches chirality, U = 0.  Taking also into account that we only consider 2 states of distinct chirality on each QD,  the on-site Coulomb interactions term in Eq.(\ref{eq:2HoleH}) reduces to \cite{rego_hawrylak_prb1997}
\begin{equation}
\label{eq:2HoleCoulomb}
U =  \frac{1}{2} \left(U_c + U_x \right) \hat{n}_{j}\hat{n}_{j'},
\end{equation}
where $\hat{n}_{j(j')}$ is the number operator, and $j,j'$ represents the 2 chirality states on a QD.  Different from electron spins, the on-site Coulomb interaction between valence holes is composed of a direct Coulomb term $U_c$ and an exchange term $U_x$ which enhance the overall interaction.  However, apart from this difference, once we have parameterized the tunneling parameter t and U, we have the Hubbard Hamiltonian.  In the strong Coulomb regime, the interaction between 2 localized particles in the Hubbard model can be reduced to two spins with the Heisenberg exchange constants $J$ proportional to $t^2/U$.  

An interesting phenomenon\cite{doty_climente_prl2009} associated with valence holes is the possibility of engineering $t$  to be either positive, zero or negative in a stack of vertically coupled cylindrical QDs by simply tuning the inter-dot distance. As the exchange coupling between 2 qubits is directly proportional to $t^2$, this implies qubit-qubit interactions can be turned on and off as needed.  We would like to investigate whether the similar phenomenon will happen with square QDs stacked either in a vertical or lateral structure.  

To estimate the magnitude of $t$,  we rely on a tight-binding picture in which a valence hole tunnels between the two dots and the hybridization of local orbitals gives a symmetric state, $1/\sqrt{2} (\vert 1 \rangle + \vert 2 \rangle)$ with energy $E_0 - t$ and an antisymmetric state, $1/\sqrt{2} (\vert 1 \rangle - \vert 2 \rangle)$ with energy $E_0 + t$.  Thus, the energy gap between first excited state and ground state is $2t$.   So we compute the energy spectrum of a single valence hole in a double QDs by exact diagonalizing Eq.(\ref{eq:LKmatrix}), then we extract the value of $t$ from the calculated energy gap.

\subsection{ Hole states in vertically coupled quantum dots}
The potential $V(\mathbf{r})$ of VCQDs  is modeled with an infinite potential well along the $x$ and the $y$ direction and double well potential profile along the $z$ direction.  In the barrier region between the dots, we set a constant finite potential of 320 $meV$ corresponding to the band offset between InGaAs and GaAs, same as reported in Ref. \onlinecite{climente_korkusinski_prb2008}, so we may draw a comparison between square-like and disk-like dots.

\begin{figure}
\centering
\includegraphics[width=0.9\textwidth]{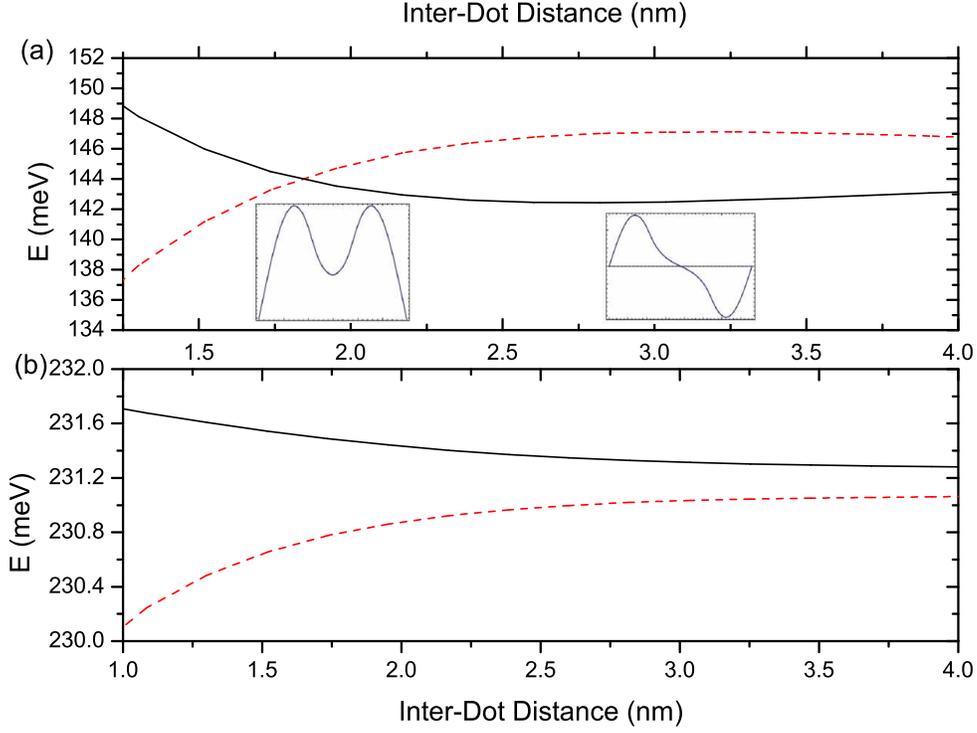}  
\caption{Energy levels of the two lowest hole states as a function of the inter-dot distance in: (a) VCQDs (vertically coupled quantum dots) and (b) LCQDs (laterally coupled quantum dots).  The critical distance at which the tunneling element is zero occurs around 1.8 nm in (a). The solid curve represents $\chi_{z}=\uparrow$ state and the dashed curve represents the $\chi_z = \downarrow$ state.  Inset: The dominant heavy hole component of the ground state before and after the crossing.}
\label{fig:cqds}
\end{figure}

As shown in Fig \ref{fig:cqds}a, the lowest energy levels of different chirality subspaces cross at a certain inter-dot distance.  The insets of Fig \ref{fig:cqds}a present the most dominant HH wavefunction profile along the $z$ direction of the ground state before and after the crossing point. The wavefunction plots show the reversal of symmetry for the ground state of the system. Similar phenomenon occurs in both square and disk dots.  These results confirm that the effects of HH and LH mixing is insensitive to the confining potential in the $x-y$ plane. These results also imply that  we can shut down any undesirable hole - hole interaction between holes localized in different dots if the inter-dot distance is chosen at the point where the crossing occurs in Fig \ref{fig:cqds}a. 

\subsection{ Hole states in laterally coupled quantum dots}
The potential of LCQDs  is modeled similar to that of VCQDs except that the double well potential profile lies along the $y$ direction.  Our result, Fig \ref{fig:cqds}b, shows no crossing of lowest energy states from distinct chirality subspaces.  We understand that the confinement strength along the $z$ direction plays a crucial role in determining how much HH and LH bands mixing a system will experience, because QDs are quasi-two dimensional devices with a much smaller dimension along the $z$ direction.  In the case of VCQDs, the coupled dots relax significantly the confinement  strength of the hole states along the $z$ direction and bring the LH and HH energies closer. However, a LCQDs structure relaxes confinement strength in the $x-y$ plane which does not facilitate the mixing of HH and LH states.  Hence no reversal of the ground state as a function of the distance is observed.

\section{Conclusion}
In summary, we consider the Luttinger spinor description of confined valence hole states in QDs.  We identify the two qubit levels with two chiralities of the lowest energy Kramers doublet and suggest that $B_z$ and $B_x$ fields act analogously to the $\hat{\chi}_z$ and $\hat{\chi}_x$ operators. For arrays of hole qubits  we study tunneling of holes in vertical and lateral architecture as dominant mechanism for the qubit exchange interaction. We show that tunneling  can be arrested for vertical pairs of quantum dots but not for lateral architecture.  The capability to switch the sign of the effective tunneling $t$ in VCQDs is demonstrated.  This suggests the possibility of turning off the exchange interaction.  With such exchange interaction being very small one can envisage tuning this interaction with additional metallic gates.

\begin{acknowledgments}
The authors thank NSERC, NRC-CNRS-CRP, QuantumWorks and CIFAR for support.
\end{acknowledgments}

\appendix*
\section{On-site Coulomb matrix elements}
The on-site Coulomb matrix elements among valence hole states $\vert j_i \rangle = \vert \chi_{z_i} \chi_{xy_i}, N_i \rangle$ can be evaluated as follows
\begin{widetext}
\begin{multline}
\label{eq:Coulomb}
U_{j_1j_2j_3j_4} =  \frac{e^2}{8 \pi^3 \epsilon_o}\sum_{J_{z_1},J_{z_2}}\sum_{\substack{n_i,m_i,l_i \\ i=1,2,3,4}}C^{*\chi_{z_1},\chi_{xy_1},N_1,J_{z_1}}_{n_1,m_1,l_1} C^{*\chi_{z_2},\chi_{xy_2},N_2,J_{z_2}}_{n_2,m_2,l_2} C^{\chi_{z_3},\chi_{xy_3},N_3,J_{z_2}}_{n_3,m_3,l_3}C^{\chi_{z_4},\chi_{xy_4},N_4,J_{z_1}}_{n_4,m_4,l_4} \\ 
\iiint d^3q \frac{1}{q^2} G_{n_1n_2 n_3n_4}(x,x',q_x) G_{m_1m_2 m_3m_4}(y,y',q_y) G_{l_1l_2 l_3l_4}(z,z',q_z),
\end{multline}
where $\epsilon_0$ is the static dielectric constants of InGaAs.  We define
\begin{equation}
G_{s_1s_2s_3s_4}(r,r',q_r)= \iint dr dr' \xi_{s_1}(r) \xi_{s_4}(r)\xi_{s_2}(r')\xi_{s_3}(r')e^{iq_{r}|r-r'|}.
\end{equation}
\end{widetext}


\newpage

\end{document}